\documentclass[11pt]{article}
\usepackage{epsf} 
\usepackage{amsmath}
\usepackage{amssymb}
\usepackage{epsfig}
\usepackage{latexsym}
\usepackage{amsfonts}
\usepackage{graphicx}
\usepackage{cite}
\usepackage{ifthen}
\usepackage{color}

\def\lsim{\:\raisebox{-0.5ex}{$\stackrel{\textstyle<}{\sim}$}\:}
\def\gsim{\:\raisebox{-0.5ex}{$\stackrel{\textstyle>}{\sim}$}\:}
\begin{document}
\parindent 0mm 
\setlength{\parskip}{\baselineskip} 
\thispagestyle{empty}
\pagenumbering{arabic} 
\setcounter{page}{0}

\hfill UCT-TP-302/14, MITP/14-074
\newline
\mbox{} \hfill Revised: February 2015
\newline
\vspace{0.1cm}
\begin{center}
{\Large {\bf 
Chiral sum rules and vacuum condensates 
\\[2mm]
from tau-lepton decay data}}
\\
\end{center}
\vspace{.05cm}

\begin{center}
{\bf  C. A. Dominguez}$^{(a),(b)}$, 
{\bf L. A. Hernandez}$^{(a)}$, 
{\bf K. Schilcher}$^{(a),(b),(c)}$, \\ {\bf and}
{\bf H. Spiesberger}$^{(a),(c)}$
\end{center}

\begin{center}
{\it $^{(a)}$Centre for Theoretical and Mathematical Physics, 
and Department of Physics, University of
Cape Town, Rondebosch 7700, South Africa}
\\

{\it $^{(b)}$National Insitute for Theoretical Physics, Private Bag X1,
Matieland 07602, South Africa}
\\

{\it $^{(c)}$ PRISMA Cluster of Excellence, Institut f\"{u}r Physik, 
Johannes Gutenberg-Universit\"{a}t, D-55099 Mainz, Germany}
\end{center}
\vspace{0.1cm}

\begin{center}
\footnotesize
{\it E-mail:} 
cesareo.dominguez@uct.ac.za, 
HRNLUI001@myuct.ac.za, 
\\
karl.schilcher@uni-mainz.de, 
spiesber@uni-mainz.de
\end{center}

\begin{center}
\textbf{Abstract}
\end{center}

QCD finite energy sum rules, together with the latest updated ALEPH 
data on hadronic decays of the tau-lepton are used in order to 
determine the vacuum condensates of dimension $d=2$ and $d=4$. 
These data are also used to check the validity of the Weinberg 
sum rules, and to determine the chiral condensates of dimension 
$d=6$ and $d=8$, as well as the chiral correlator at zero momentum, 
proportional to the counter term of the ${\cal{O}}(p^4)$ Lagrangian 
of chiral perturbation theory, $\bar{L}_{10}$. Suitable (pinched) integration 
kernels are introduced in the sum rules in order to suppress 
potential quark-hadron duality violations. We find no compelling 
indications of duality violations in the kinematic region above 
$s \simeq 2.2$ GeV$^2$ after using pinched integration kernels. 

%{\color{red}
%TO BE DONE: 
%\\
%Check all numbers, in particular for C4O4.
%}
%
%{\color{blue} Date: \today}

\clearpage

%%%%%%%%%%%%%%%%%%%%%%%%%%%%%%%%%%%%%%%%%%%%%%%%%%%%%%%%%%%%%%%%%%%%%%%%%

\section{Introduction}

Experimental data on hadronic decays of the $\tau$-lepton 
\cite{ARGUS,ALEPH1,ALEPH1new,ALEPH2} play an essential role in 
the determination of several QCD quantities \cite{PICH1}. For 
instance, the $R_\tau$-ratio provides the cleanest determination 
of the running strong coupling at the scale of the $\tau$-mass. 
In addition, these data have been used in the past to extract 
the values of some of the QCD vacuum condensates entering the 
operator product expansion (OPE) of current correlators at short 
distances beyond perturbation theory \cite{COND1,COND2,ALPHA,COND3, 
Almasy,COND3b,COND4}. This OPE is one of the two fundamental pillars 
of the method of QCD sum rules, the other being the assumption of 
quark-hadron duality \cite{QCDReview}. The latter allows to relate 
QCD with hadronic physics by means of Cauchy's theorem in the 
complex squared energy plane. A key advantage of hadronic 
$\tau$-decay data is that it determines both the vector and the 
axial-vector spectral functions. This feature  allows to check 
the saturation of a variety of chiral sum rules \cite{COND3}, 
\cite{CHSR1,GMOR,CHSR2}, as well as to determine the chiral 
correlator at zero momentum \cite{PICH1}, \cite{CHSR2,L101,DMO, 
L102,KSMT,L10F}, proportional to the counter term of the order 
${\cal{O}}(p^4)$ Lagrangian of chiral perturbation theory (CHPT), 
$\bar{L}_{10}$. It also allows for a determination of the chiral 
condensates of dimension $d=6$ and $d=8$ \cite{PICH1,COND3,CHSR2, 
L101,DMO,L102,KSMT,L10F}.

Most of these past determinations made use of the hadronic 
spectral functions in the vector and axial-vector channel as 
measured by the ALEPH Collaboration \cite{ALEPH1,ALEPH1new}. 
This data base was known to be problematic due to the incompleteness 
of the data correlations \cite{Boito}, thus casting some doubt on 
the uncertainties in  results obtained using these data. A new 
ALEPH data set has recently become available \cite{ALEPH2}, with 
the data organized in different bins, and with a corrected error 
correlation matrix. In this paper we employ these data to revisit 
the vacuum condensate determinations, the saturation of chiral 
sum rules, and the determination of $\bar{L}_{10}$ and the chiral 
condensates of dimension $d=6$ and $d=8$. The procedure is based 
on finite energy QCD sum rules (FESR), weighted with suitable 
integration kernels to account for potential duality violations (DV). 
Our results mostly confirm central values obtained previously 
using the original ALEPH data base, with uncertainties being 
slightly higher in some cases, and lower in others.

%%%%%%%%%%%%%%%%%%%%%%%%%%%%%%%%%%%%%%%%%%%%%%%%%%%%%%%%%%%%%%%%%%%%%%%%%

\section{QCD finite energy sum rules and vacuum condensates}

We consider the (charged) vector and axial-vector current 
correlators
%Eq.1
\begin{align}
\Pi_{\mu\nu}^{VV}(q^{2})  
&= 
i\int d^{4}x \; e^{iqx} 
\langle 0|T(V_{\mu}(x)V_{\nu}^{\dagger}(0))|0\rangle 
\label{Eq:1} 
\\
&= 
(-g_{\mu\nu}\;q^{2}+q_{\mu}q_{\nu})\;\Pi_{V}(q^{2}) \; , 
\nonumber
\end{align}
%Eq.2
\begin{align}
\Pi_{\mu\nu}^{AA}(q^{2})  
&= 
i\int d^{4}x \;e^{iqx} 
\langle 0|T(A_{\mu}(x)A_{\nu}^{\dagger}(0))|0\rangle 
\label{Eq:2} 
\\
&= \; 
(-g_{\mu\nu}q^{2}+q_{\mu}q_{\nu}) \; 
\Pi_{A}(q^{2})-q_{\mu}q_{\nu} \; 
\Pi_{0}(q^{2}) \; , 
\nonumber
\end{align}
where $V_{\mu}(x) = :\bar{u}(x)\gamma_{\mu}d(x):$, $A_{\mu}(x) = 
:\bar{u}(x)\gamma_{\mu}\gamma_{5}d(x):$, with $u(x)$ and $d(x)$ 
the quark fields, and $\Pi_{V,A}(q^{2})$ normalized in perturbative 
QCD (PQCD) (in the chiral limit) according to 
%Eq.3
\begin{equation}
\frac{1}{\pi}\operatorname{Im}\Pi_{V}^{PQCD}\left(s\right) 
= \frac{1}{\pi}\operatorname{Im}\Pi_{A}^{PQCD} 
\left(s\right) = 
\frac{1}{4\pi^{2}}\left(1+\frac{\alpha_{s}(s)}{\pi} +...\right) \;,  
\label{Eq:3}
\end{equation}
where $s \equiv q^2 > 0$ is the squared energy. Lorentz decomposition 
is used to separate the correlation function into its $J=1$ and $J=0$ 
parts. To the accuracy needed in the following, the vector current 
can be assumed to be conserved. The correlators are well-known up 
to five-loop order \cite{CHET1}. Solving the renormalization group 
equation for the strong coupling, one can express $\alpha_s(s)$ in 
terms of the coupling at a given scale $s_0$, with the result at 
six-loop order being \cite{ALPHA}
%Eq.4
\begin{eqnarray}
a_s(s) 
&=& 
a_s(s_0) 
+ a_s^2(s_0) \, \left(\frac{1}{2}\, \beta_1 \, \eta\right) 
+ a_s^3(s_0) \left(\frac{1}{2}\, \beta_2 \, \eta + \frac{1}{4} \, 
\beta_1^2 \,\eta^2\right) 
\nonumber 
\\[.4cm]
& + & 
a_s^4(s_0) \left[\frac{1}{2} \, \beta_3 \, \eta + \frac{5}{8} \, 
\beta_1 \,\beta_2 \, \eta^2 + \frac{1}{8} \, \beta_1^3 \, 
\eta^3\right]
\nonumber 
\\[.4cm]
& + & 
a_s^5(s_0) \left[- b_3 \, \eta + \frac{3}{8} \, \beta_2^2  \, \eta^2 
+ \frac{3}{4} \, \beta_1 \, \beta_3 \, \, \eta^2 
+ \frac{13}{24} \, \beta_1^2 \, \beta_2 \, \eta^3 
+ \frac{1}{16} \, \beta_1^4 \, \eta^4  \right]
\label{Eq:4}
\end{eqnarray}
with 
%Eq.5
\begin{equation}
\eta = \ln\left(\frac{s}{s_0}\right) \, .
\label{Eq:5}
\end{equation}
The coefficients of the $\beta$-function are given by
%Eq.6
\begin{eqnarray}
\beta_1 &=& - \frac{1}{2} \left(11 - \frac{2}{3} n_F\right) \, , 
\quad 
\beta_2 \, = \, - \frac{1}{8} \left(102 - \frac{38}{3} n_F \right) \, , 
\nonumber
\\
\beta_3 &=& - \frac{1}{32} 
\left(\frac{2857}{2} - \frac{5033}{18} n_F + \frac{}{} n_F^2\right) 
\, ,
\label{Eq:5a}
\end{eqnarray}
and
%Eq.7
\begin{eqnarray}
b_3 
& = & 
\frac{1}{4^4} \Biggl[ \frac{149753}{6} + 3564 \zeta_3
- \left(\frac{1078361}{162} + \frac{6508}{27} \zeta_3 \right) n_F 
\nonumber 
\\
& + & 
\left(\frac{50065}{162} + \frac{6472}{81} \zeta_3 \right) n_F^2 
+ \frac{1093}{729} n_F^3 \Biggr ] \; , 
\label{Eq:6}
\end{eqnarray}
with $\zeta_{3} = 1.202$. 

Non-perturbative contributions are parametrized in terms of the 
vacuum condensates entering the OPE
%Eq.8
\begin{equation}
4\pi^2 \Pi(Q^{2})|_{V,A} 
= 
\sum_{N=1}^{\infty}\frac{1}{Q^{2N}} \; 
C_{2N}(Q^{2},\mu^{2})\; 
\langle O_{2N}(\mu^{2})\rangle|_{V,A}\;, 
\label{Eq:7}
\end{equation}
where $Q^2 = -q^2$, and $\mu$ is a renormalization scale separating 
long distance non-perturbative effects associated with the vacuum 
condensates $\langle O_{2N}(\mu^{2})\rangle$ from the short distance 
effects which are encapsulated in the Wilson coefficients 
$C_{2N}(Q^{2},\mu^{2})$. In principle, the lowest dimension is 
$d=4$ as there are no gauge invariant operators of dimension $d=2$ 
in QCD. However, the absence of such a condensate will be confirmed 
by the results of this analysis. At dimension $d=4$, and in the 
chiral limit, the only contribution is from the (chiral-symmetric) 
gluon condensate
%Eq.9
\begin{equation}
C_4 \langle O_4 \rangle|_{V,A} 
= 
\frac{\pi^2}{3}\, 
\langle \frac{\alpha_s}{\pi}\, G_{\mu\nu}\, G^{\mu\nu} \rangle 
\;,
\label{Eq:8}
\end{equation}
where $\alpha_s$ is the running strong coupling, and in the 
sequel $\langle 0| O_{2N} |0 \rangle \equiv \langle O_{2N} \rangle$ 
is to be understood. This condensate is renormalization group 
invariant to all orders in PQCD (in the chiral limit).
\\

Invoking Cauchy's theorem in the complex squared energy $s$-plane, 
and assuming (global) quark-hadron duality leads to the FESR
%Eq.10
\begin{equation}
- \frac{1}{2 \pi i} \; \oint_{|s|=s_{0}} ds\; f(s) \; 
\Pi(s)|^{QCD}_{V,A} \; 
= 
\int_{0}^{s_{0}} ds \; f(s) \; 
\rho_{V,A}(s) \; ,
\label{Eq:9}
\end{equation} 
where $f(s)$ is an integration kernel and $\rho_{V,A}(s)$ are the 
hadronic spectral functions, 
%Eq.11
\begin{equation}
\rho_{V,A}(s) 
= 
\frac{1}{\pi}\; {\mbox{Im}}\; \Pi(s)|^{HAD}_{V,A} 
= 
\frac{1}{2\pi^2} \left[v(s), a(s)\right]_{\mbox{\footnotesize ALEPH}} 
\label{Eq:rhova}
\end{equation}
provided by the ALEPH data. Since PQCD is not applicable 
on the positive real $s$-axis, a very early warning against the 
unqualified use of sum rules was raised \cite{Shankar} even before 
the QCD sum rule program was proposed. A priori it is not clear 
at which scale duality sets in. It was shown \cite{CHSR1,PINCH1} 
that by reducing the impact of $\Pi(s)|_{V,A}^{QCD}$ in the 
contribution of the integration contour near the positive real axis 
in Eq.\ (\ref{Eq:9}) by a suitable integration kernel $f(s)$ 
(pinching), the range of manifest duality can be increased 
substantially. In particular, we have shown previously for the 
old ALEPH data that there is clear evidence that duality is satisfied 
towards the end of the decay spectrum\cite{COND4}. In practice, the 
absence of (DV) can be inferred from sum rules where 
their values are known from other sources or, less compelling, from 
the stability of the integral against variations of the upper limit 
of integration $s_{0}$. We will demonstrate below that duality can 
be observed with the new ALEPH data for many sum rules. However,  DV is a contentious issue relying on specific models \cite{L101},\cite{L102}, as discussed in more detail in Section 3.

The contour integral in Eq.\ (\ref{Eq:9}) is usually computed 
using fixed order perturbation theory (FOPT) or contour improved 
perturbation theory (CIPT). In the former case the strong coupling 
is frozen at a scale $s_0$ and the renormalization group (RG) is 
implemented after integration. In CIPT $\alpha_s(s)$ is running 
and the RG is used before integrating, thus requiring solving 
numerically the RG equation for $\alpha_s(s)$ at each point on 
the integration contour. In the specific case of the determination 
of the vacuum condensates we found CIPT to be superior to FOPT in 
that results turn out to be more stable as a function of $s_0$. 
To implement CIPT it is convenient to introduce the Adler function
%Eq.12
\begin{equation}
D(s) \equiv -s \frac{d}{ds} \Pi(s) \; , 
\label{Eq:10}
\end{equation}
with $\Pi(s) \equiv \Pi_{V,A}(s)$. Invoking Cauchy's theorem and 
after integration by parts the following relation is obtained
%Eq.13
\begin{equation}
\oint_{|s|=s_{0}} ds\; \left(\frac{s}{s_{0}}\right)^{N} \;\Pi(s) 
= 
\frac{1}{N+1} \; 
\frac{1}{s_{0}^{N}} \;\oint_{|s|=s_{0}} \frac{ds}{s} \; 
\left(s^{N+1} - s_{0}^{N+1}\right)\; D(s) \; . 
\label{Eq:11}
\end{equation}
After RG improvement, the perturbative expansion of the Adler 
function becomes
%Eq.14
\begin{equation}
D(s) 
= 
\frac{1}{4\; \pi^{2}} \sum_{m = 0} K_{m} \; 
\Bigl[\frac{\alpha_{s}(-s)}{\pi}\Bigr]^{m} \; , 
\label{Eq:12}
\end{equation}
where \cite{CHET1} $K_{0} = K_{1} =1$, $K_{2} = 1.6398$ , 
$K_{3} = 6.3710$, for three flavours, and  $K_{4} = 49.076$ 
\cite{K4}. The vacuum condensates are then determined from the 
pinched FESR
%Eq.15
\begin{align}
C_{2N+2} \langle O_{2N+2} \rangle 
&= 
(-)^{N+1}\; 4 \pi^{2}\; s_{0}^{N} \;
\int_{0}^{s_{0}} \; ds 
\left[  1 - \left(\frac{s}{s_{0}}\right)^{N} \right]  
\frac{1}{\pi} {\mbox{Im}} \; \Pi(s)^{HAD} 
\nonumber 
\\[.4cm]
& + (-)^{N} s_{0}^{N+1} [ M_{0}(s_{0}) - M_{N}(s_{0})] \;, 
\label{Eq:13}
\end{align}
where the moments $M_N(s_0)$ are given by
%Eq.16
\begin{equation}
M_{N}(s_{0}) 
= 
\frac{1}{2 \pi} \frac{1}{(N+1)} \sum_{m=0} K_{m} \;
[I_{N+1,m}(s_{0}) - I_{0,m}(s_{0})] \;, 
\label{Eq:14}
\end{equation}
with
%Eq.17
\begin{equation}
I_{N,m} \equiv i \; \oint_{|s|=s_{0}} ds\; 
\left(\frac{s}{s_{0}}\right)^{N} \;
\left[\frac{\alpha_{s}(-s)}{\pi}\right]^{m} \; . 
\label{Eq:15}
\end{equation}

The latest ALEPH data compilation \cite{ALEPH2} includes the 
vector and axial-vector channels separately, as well as their 
sum. Their data are given in tables for the normalised invariant 
mass-squared distributions. We determine the spectral functions 
as described, for example, in \cite{ALEPH1new} and approximate 
the sum rule integrals by sums over bins, taking into account 
the corrected correlation matrix. We should note that we will 
omit the last two points with the highest $s$ values in the 
figures for our results to be discussed below; they have very 
large experimental uncertainties and do not affect our conclusions.  
We use the following values 
for the input parameters 
%Eq.18
\begin{align}
m_{\pi} &= 139.57018(35)~\mbox{MeV} \, ,
\label{1.1} \\
f_{\pi} &= 92.21(14)~\mbox{MeV} \, , 
\nonumber\\
M_{\tau} &= 1776.82(16)~\mbox{MeV} \, , 
\nonumber\\
V_{ud} &= 0.97425(22) \, ,
\nonumber
\end{align}
\begin{equation}
S_{EW} = 1.0198\, ,
\quad 
B_{e} = 0.17818 \, .
\nonumber
\end{equation}
The first four values are taken from the particle data group \cite{PDG}. 
$S_{EW}$ is needed to include the renormalization-group improved 
electroweak corrections \cite{MarSir}. As the leptonic branching 
ratio $B_{e}$ was not updated in the recent paper \cite{ALEPH1new}, 
we again use the one given in the earlier ALEPH report \cite{ALEPH1}. 
From the latest analysis \cite{PICH1}, we have $\alpha_{s}(M_{\tau}^{2}) 
= 0.341 \pm 0.013$ in CIPT and $\alpha_{s}(M_{\tau}^{2}) = 0.319 \pm 
0.014$ in FOPT. For consistency we use the CIPT result in the 
following.  

%Fig.1

%%%%%%%%%%%%%%%%%%%%%%%%%%%%%%%%%%%%%%%
\begin{figure}[t!]
\begin{center}
\includegraphics[width=5.2in]{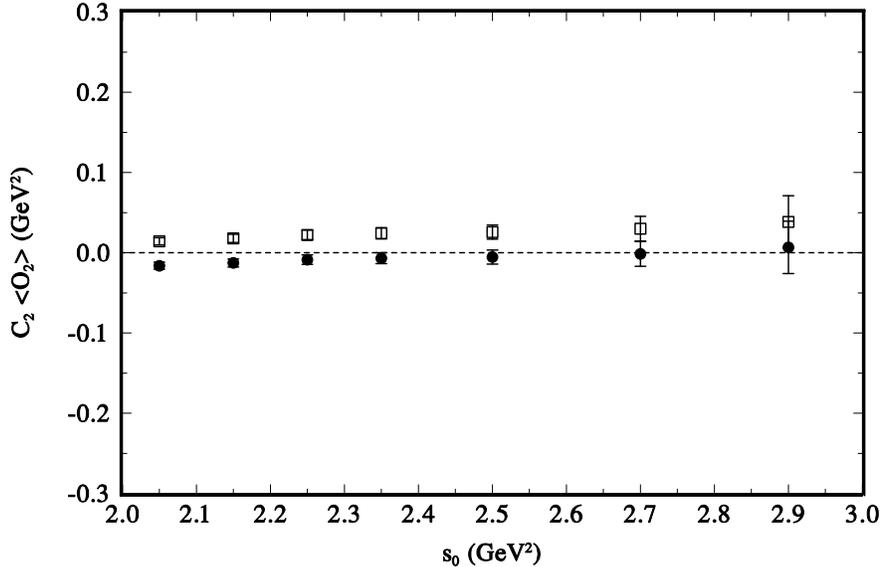}
\caption{
\footnotesize{
The dimension $d=2$ condensate in CIPT from the  FESR, Eq.\ 
(\ref{Eq:13}), with $N=0$. The ALEPH data for the $V+A$ spectral 
function was used, and the resulting condensate divided by 2. 
The two sets of points correspond to $\alpha_s = 0.354$ (full dots) 
and $\alpha_s = 0.328$ (open squares). 
}
}
\label{Fig:d=2}
\end{center}
\end{figure} 
%%%%%%%%%%%%%%%%%%%%%%%%%%%%%%%%%%%%%%%

Proceeding with the determination of a potential $d=2$ 
condensate (presumably chiral-symmetric) we have used the data 
base for $V+A$ in the FESR and divided the answer by a factor two. 
In Fig.\ \ref{Fig:d=2} we show the result in the stability region. 
The solid dots correspond to the minimum value of $\alpha_s$ and 
the open squares to its maximum value. As expected, this $d=2$ 
term is consistent with zero. Notice that in this case there is 
no pinching integration kernel as $N=0$ in Eq.\ (\ref{Eq:13}). 

%Fig.2
%%%%%%%%%%%%%%%%%%%%%%%%%%%%%%%%%%%%%%%
\begin{figure}[t]
\begin{center}
\includegraphics[width=5.2in]{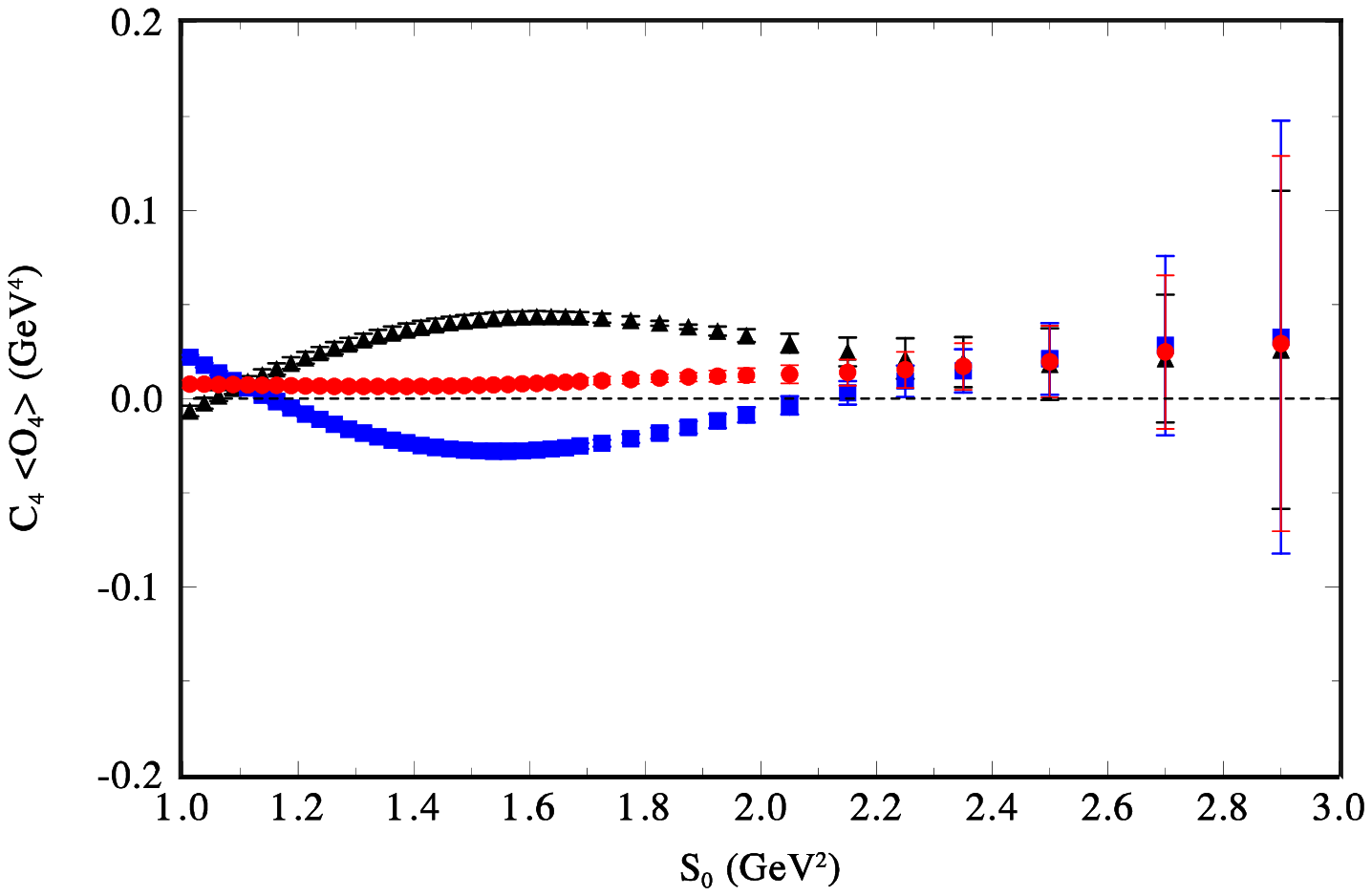}
\caption{
\footnotesize{
The dimension $d=4$ condensate in CIPT from the FESR, Eq.\ 
(\ref{Eq:13}), with $N=1$ and $\alpha_s(M_\tau^2) = 0.341$. 
The ALEPH data for the $V$ (upper black dots), $A$ (lower blue dots) 
and $\frac{1}{2}(V+A)$ (middle red dots) spectral function were used. 
}
}
\label{Fig:d=4}
\end{center}
\end{figure} 
%%%%%%%%%%%%%%%%%%%%%%%%%%%%%%%%%%%%%%%

Next, we make use of this result and consider the pinched FESR, 
Eq.\ (\ref{Eq:13}), with $N=1$. The condensate of $d = 4$ is shown 
in Fig.\ \ref{Fig:d=4}, for $V$, $A$ and $\frac{1}{2}(V+A)$. We 
observe that for $s_0 \gsim 2.2$ GeV$^2$ and within errors 
%Eq.19
\begin{equation}
C_{4}\langle O_{4} \rangle_{V} 
= 
C_{4}\langle O_{4} \rangle_{A} 
= 
C_{4}\langle O_{4}\rangle_{\frac{1}{2}(V+A)} 
\label{Eq:O4VA}
\end{equation}
over a wide range of $s_{0}$. This equality is an essential result of QCD. In addition the $d=4$ condensate is generally expected 
to be positive because it is dominated by the gluon condensate, 
Eq.\ (\ref{Eq:8}), which in turn is directly related to the vacuum 
energy density \cite{SVZ}, 
%Eq.20
\begin{equation}
\varepsilon 
= 
\frac{\pi}{8\alpha_{s}^{2}} \beta(\alpha_{s}) 
\langle\frac{\alpha_{s}}{\pi}\,G_{\mu\nu}\,G^{\mu\nu}\rangle \, .
\end{equation}
Therefore, the sign and magnitude of the gluon condensate 
$\langle\frac{\alpha_{s}}{\pi}\,G_{\mu\nu}\,G^{\mu\nu}\rangle$ are 
of fundamental importance for the understanding of the strong 
interactions. A negative value of $\varepsilon$ is expected 
from models such as the bag model. In our analysis we obtain for 
$\alpha_s(M_\tau^2) = 0.341$  
%Eq.21
\begin{equation}
C_{4}\langle O_{4}\rangle 
= 
(0.017 \pm 0.012) \; {\mbox{GeV}}^{4} \; ,
\label{Eq:16}
\end{equation}
where this value is obtained by reading results from the $V+A$ 
spectral function at $s_0 = 2.35$ GeV$^2$, i.e.\ at the point 
where $C_{4}\langle O_{4}\rangle$ from the $V$ and $A$ channels 
become equal. This value is consistent within errors with the 
points at higher values of $s_0$ and agrees with a previous 
determination \cite{COND4} using the original ALEPH data base 
\cite{ALEPH1}. However, the uncertainty is now larger due to the 
new ALEPH error correlation matrix. 
%If one were to use the data 
%separately in the vector and the axial-vector channel, then the 
%result from the former would be $C_{4} \langle  O_{4} \rangle = 
%(0.021 \pm 0.011)$ GeV$^4$ and $C_{4} \langle O_{4} \rangle = 
%(0.009 \pm 0.008)$ GeV$^4$ from the latter. However, the 
%axial-vector data on its own is not particularly stable, thus leading 
%to a result with a larger uncertainty. This behaviour is understood 
%from the fact that the $\rho(770)$-meson is a narrow resonance at a 
%lower energy than the much broader $a_1(1260)$, leading to a different 
%saturation of the hadronic integrals in the FESR. 
We observe that the precise value for $\alpha_s$ chosen in the 
evaluation of the condensate has a relatively large impact on 
the result: the uncertainty of $\pm 0.013$ for $\alpha_s(M_\tau^2)$ 
given in \cite{PICH1} gives rise to an additional uncertainty of 
$\pm 0.018$ for $C_{4} \langle O_{4} \rangle$. We repeated the 
analysis using FOPT. The results are very similar, though. For 
example, for the central FOPT value $\alpha_s(M_\tau^2) = 0.319$ 
we obtain $C_{4} \langle O_{4} \rangle = (0.022 \pm 0.006)$ 
GeV$^4$. Combining results, we can say that all evidence points 
to a positive value of $C_{4} \langle O_{4} \rangle \lsim 0.035$ 
GeV$^4$ which is equal for the vector and the axial-vector 
correlators. In contrast, in the updated analysis of the ALEPH 
data \cite{ALEPH2} unequal and negative results for the $V$ and 
$A$ channels have been obtained. 

%Fig.3
%%%%%%%%%%%%%%%%%%%%%%%%%%%%%%%%%%%%%%%
\begin{figure}[h!]
\begin{center}
\includegraphics[width=5.2in]{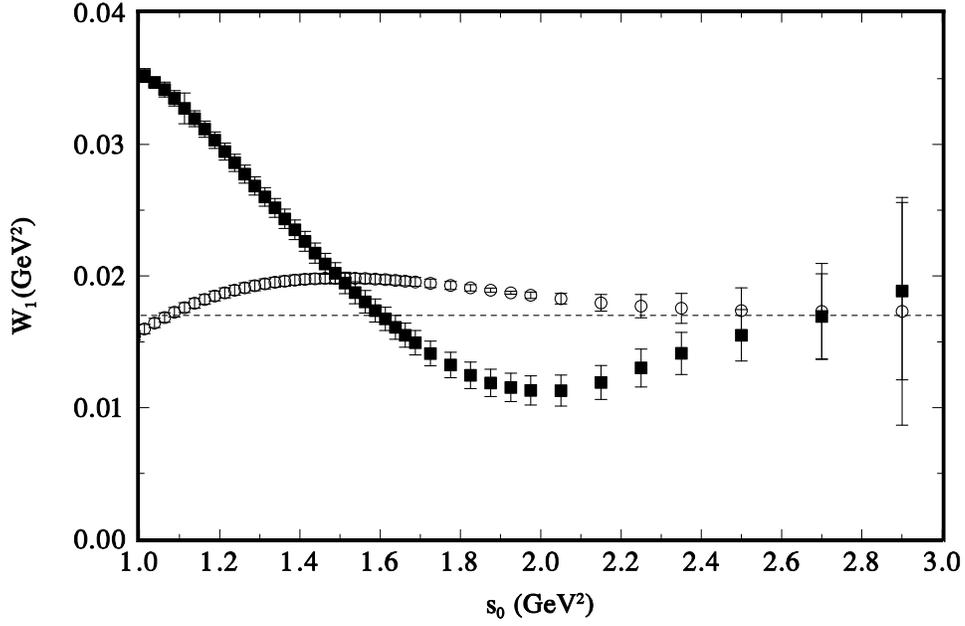}
\caption{
\footnotesize{
Solid squares are the left-hand-side of the standard  Weinberg 
sum rule, Eq.\ (\ref{Eq:17}), and open circles the left hand side 
of the pinched sum rule, Eq.\ (\ref{Eq:19}). The dotted line is 
the right-hand-side, $2 f_\pi^2$. 
}
}
\label{Fig:W1P}
\end{center}
\end{figure} 
%%%%%%%%%%%%%%%%%%%%%%%%%%%%%%%%%%%%%%%

%Fig.4
%%%%%%%%%%%%%%%%%%%%%%%%%%%%%%%%%%%%%%%
\begin{figure}[h!]
\begin{center}
\includegraphics[width=5.2in]{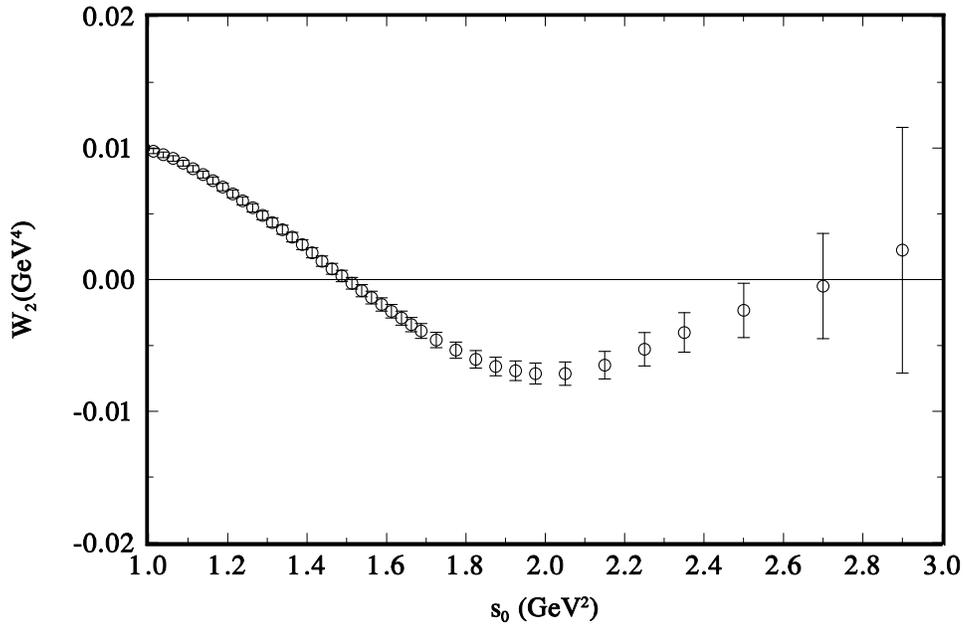}
\caption{
\footnotesize{
The second Weinberg sum rule $W_2$ as a function the upper 
limit of integration. 
}
}
\label{Fig:W2P}
\end{center}
\end{figure} 
%%%%%%%%%%%%%%%%%%%%%%%%%%%%%%%%%%%%%%%

The next condensates, i.e.\ with dimension $d=6$, in the vector and 
the axial-vector channels do not show a stability region. This type 
of FESR is not suited to extract higher dimensional condensates 
because the power weight in the FESR increasingly emphasizes the 
high energy region, where experimental errors are large and where 
the condensates are the result of a fine balanced cancellation 
between the hadronic integral and the PQCD moments, with a marginally 
meaningful result at $d=4$, but not beyond. In the next section we 
shall determine the chiral condensates of dimension $d=6$ and $d=8$, 
which do not suffer from this handicap as the perturbative 
contribution cancels exactly (in the chiral limit).

%%%%%%%%%%%%%%%%%%%%%%%%%%%%%%%%%%%%%%%%%%%%%%%%%%%%%%%%%%%%%%%%%%%%%%%%%

\section{Chiral sum rules and chiral vacuum condensates}

The two Weinberg sum rules (WSR) \cite{WSR} were first derived in 
the framework of chiral $SU(2) \times SU(2)$ symmetry and current 
algebra, retaining their validity in QCD in the chiral limit, and 
read
%Eq.22
\begin{equation}
W_1 
\equiv 
\int\limits_{0}^{\infty} ds \,\frac{1}{\pi}\, 
[{\mbox{Im}} \Pi_V(s) - {\mbox{Im}} \Pi_A(s) ] = 2 \, f_\pi^2\;,
\label{Eq:17}
\end{equation}
%Eq.23
\begin{equation}
W_2 
\equiv 
\int\limits_{0}^{\infty} ds \,s \,\frac{1}{\pi}\, 
[{\mbox{Im}} \Pi_V(s) - {\mbox{Im}} \Pi_A(s) ] =0 \;,
\label{Eq:18}
\end{equation}
where $f_\pi = 92.21 \pm 0.14\, {\mbox{MeV}}$ \cite{PDG}. The 
integration region can be split into two parts, one in the range 
$0-s_0$ and the other in $s_0-\infty$. Since the spectral function 
difference vanishes in PQCD for $s_0$ sufficiently large, these 
sum rules effectively become FESR. However, as pointed out long ago \cite{CHSR1,CHSR2}, the original 
$\tau$-decay ALEPH data \cite{ALEPH1} did not saturate these 
integrals up to the kinematic end point $s_0 \simeq M_{\tau}^2$.
This could also be said for the updated ALEPH data \cite{ALEPH2} 
if the existence of a plateau of the central values is taken as 
a criterion for saturation (see solid squares in Fig.\ \ref{Fig:W1P} 
for $W_1$ and open circles in Fig.\ \ref{Fig:W2P} for $W_2$). The size of the 
experimental uncertainties, however, does  not allow us to 
conclude that saturation has not been reached. A much better behaviour
is achieved  after introducing a simple pinched kernel and 
combining the two sum rules into one
%Eq.24
\begin{equation}
W_{1P}(s_0) 
\equiv
\int\limits_{0}^{s_0} ds \left( 1 - \frac{s}{s_0}\right) 
\frac{1}{\pi}\; [{\mbox{Im}} \Pi_V(s) - {\mbox{Im}} \Pi_A(s) ] \;
= 
2\,  f_\pi^2 \,.
\label{Eq:19}
\end{equation}
The result is shown in Fig.\ \ref{Fig:W1P} (open circles), indicating 
a very good saturation of the pinched sum rule. This supports the use of 
simple integration kernels, although DV  could be 
channel or application dependent. 

The two Weinberg sum rules are particularly interesting since 
they would not be satisfied if there were substantial DV  present, i.e.\ non-perturbative contributions beyond 
perturbative QCD and OPE contributions. The issue of DV  is indeed most prominent in the context of the 
$V-A$ correlator, since the perturbative component cancels out
leaving a purely non-perturbative result. The two simple, i.e.\ 
un-pinched Weinberg sum rules agree with the OPE expectations 
only near the end of the decay spectrum $s_{0} \simeq 2.7$ GeV$^{2}$. 
Because of experimental limitations, the errors are relatively 
large and no definite conclusions on the relevance of duality 
violations can be drawn in this case. The last two experimental 
points should be ignored in the discussion because they cannot be
accommodated either by PQCD and the OPE or in models for DV . The pinched sum rule, however, is saturated beginning 
at $s_{0} \geq 2.2$ GeV$^{2}$ and shows remarkable agreement with 
the prediction of $2f_{\pi}^{2}$. No compelling evidence is seen 
for the existence of DV  in this kinematic domain. 
We assert that for simple un-pinched Weinberg sum rules, possible 
DV  are not required for $s_0\gsim 2.7$ GeV$^2$ 
while for the pinched sum rule possible DV  can be 
ignored beginning at much lower momentum transfers, i.e.\ already 
for $s_0\gsim 2.2$ GeV$^2$.  In view of our result it  seems very reasonable to take over this conclusion to the separate $V$ and $A$ sum rules.  
The lack of evidence for DV  in the separate $V$ 
and $A$ correlators at large $s_0$ was also demonstrated in 
\cite{DNS-PRD80}, albeit with the old ALEPH data. These conclusions are in contrast  with those following from specific models of DV  \cite{L101},\cite{L102}.

%Fig.5
%%%%%%%%%%%%%%%%%%%%%%%%%%%%%%%%%%%%%%%
\begin{figure}[ht]
\begin{center}
\includegraphics[width=5.2in]{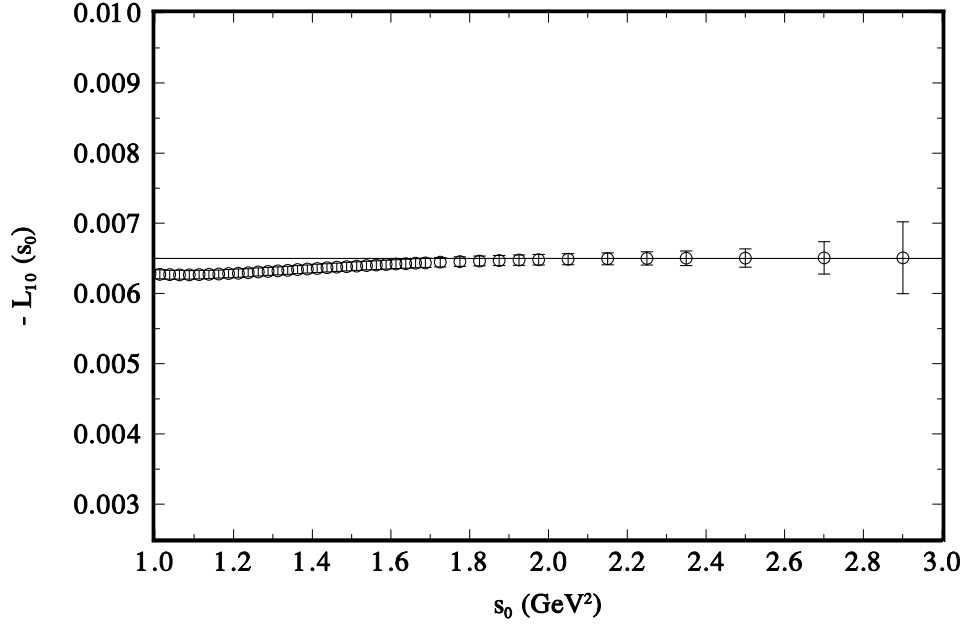}
\caption{
\footnotesize{
The CHPT constant $- \bar{L}_{10}$ obtained from the pinched chiral 
sum rule  for $\bar{\Pi}(0)$ Eq.\ (\ref{Eq:21}). 
}
}
\label{Fig:L10}
\end{center}
\end{figure} 
%%%%%%%%%%%%%%%%%%%%%%%%%%%%%%%%%%%%%%%

Next, we consider the chiral correlator $\Pi(Q^2)|_{V-A}$, and 
absorb the Wilson coefficients entering Eq.\ (\ref{Eq:7}) into 
the operators, renaming them ${\cal{O}}_N$ to conform with a 
usual convention in the literature, 
%Eq.25
\begin{equation}
\Pi(Q^{2})|_{V-A} 
= 
\sum_{N=1}^{\infty}\frac{1}{Q^{2N+4}}\; 
\langle {\cal{O}}_{2N+4} \rangle\;, 
\label{Eq:20}
\end{equation}
with the first two chiral condensates being $\langle {\cal{O}}_{6} 
\rangle$ and $\langle{\cal{O}}_{8} \rangle$. Dropping the label 
$V-A$, the finite remainder of this chiral correlator at zero 
momentum, $\bar{\Pi}(0)$, is given by
%Eq.26
\begin{equation}
\bar{\Pi}(0) 
= 
\int_0^{s_0} \, \frac{ds}{s} \, 
\frac{1}{\pi}
[ {\mbox{Im}} \Pi_V(s) - {\mbox{Im}} \Pi_A(s)] \;,
\label{Eq:21}
\end{equation} 
where ${\mbox{Im}} \Pi_A(s)$ does not include the pion pole. 

The chiral correlator at zero momentum, $\bar{\Pi}(0)$, is 
determined by the Das-Mathur-Okubo (DMO) sum rule \cite{DMO}, 
%Eq.27
\begin{equation}
\bar{\Pi}(0) 
= 
2 \left( \frac{1}{3} \; f_\pi^2 \, 
\langle r^2_\pi \rangle \,-\, \frac{1}{2} F_A \right) 
= 
0.0520 \pm 0.0010 \; ,
\label{Eq:22}
\end{equation}
where $\langle r^2_\pi \rangle = 0.439 \pm 0.008$ fm$^2$ is the 
electromagnetic radius of the pion \cite{PIONR}, and $F_A = 0.0119 
\pm 0.0001$ is the radiative pion decay constant \cite{PDG}. 
Since the numerical value on the right-hand side of Eq.\ 
(\ref{Eq:22}) is known with high precision, the DMO sum rule is 
another case where DV  would easily become visible. 
Our results in Fig.\ \ref{Fig:L10} show a wide stability region 
starting already at $s_0 \simeq 2$ GeV$^2$ for the pinched DMO 
sum rule (using Eqs.\ (\ref{Eq:17}), (\ref{Eq:18})) 
%Eq.28
\begin{equation}
\bar{\Pi}(0) 
= 
4 \,\frac{f_\pi^2}{s_0}\, + \, \int_0^{s_0} \, \frac{ds}{s} \, 
\left(1 - \frac{s}{s_0}\right)^2 
\frac{1}{\pi}
[{\mbox{Im}} \Pi_V(s) - {\mbox{Im}} \Pi_A(s)] \; . 
\label{Eq:23}
\end{equation}
$\bar{\Pi}(0)$ is proportional to the counter term of the order 
${\cal{O}}(p^4)$ Lagrangian of chiral perturbation theory, 
$\bar{L}_{10}$ \cite{ChPT-L10}, 
%Eq.27a
\begin{equation}
\bar{\Pi}(0) 
= 
- 8 \; \bar{L}_{10} \, .
\label{Eq:22a}
\end{equation}
We find 
\begin{equation}
\bar{L}_{10} = - (6.5 \pm 0.1) \times 10^{-3} \, .
\end{equation}
This result is in very good agreement with an early determination 
based on the original ALEPH data base \cite{CHSR2}, $\bar{L}_{10} 
= - (6.43 \pm 0.08) \times 10^{-3}$, as well as with more recent 
results using more involved methods to deal with DV , 
e.g. $\bar{L}_{10} = - (6.46 \pm 0.15) \times 10^{-3}$ from 
\cite{L101}, and $\bar{L}_{10} = - (6.52 \pm 0.14) \times 10^{-3}$ 
from \cite{L102}. It also agrees with lattice QCD determinations 
within their larger uncertainties \cite{LQCD}.

The relation between $\bar{\Pi}(0)$ and the precisely known 
quantities $f_\pi^2$, $\langle r^2_\pi \rangle$ and $F_A$, is 
another case where the presence of DV  can be 
tested. We observe that our result shown in Fig.\ \ref{Fig:L10} 
is very stable with respect to variations of $s_0$ in the 
range above 2 GeV$^2$ and the result from CHPT is reproduced 
with amazingly good accuracy.

%Fig.6
%%%%%%%%%%%%%%%%%%%%%%%%%%%%%%%%%%%%%%%
\begin{figure}[ht]
\begin{center}
\includegraphics[width=5.2in]{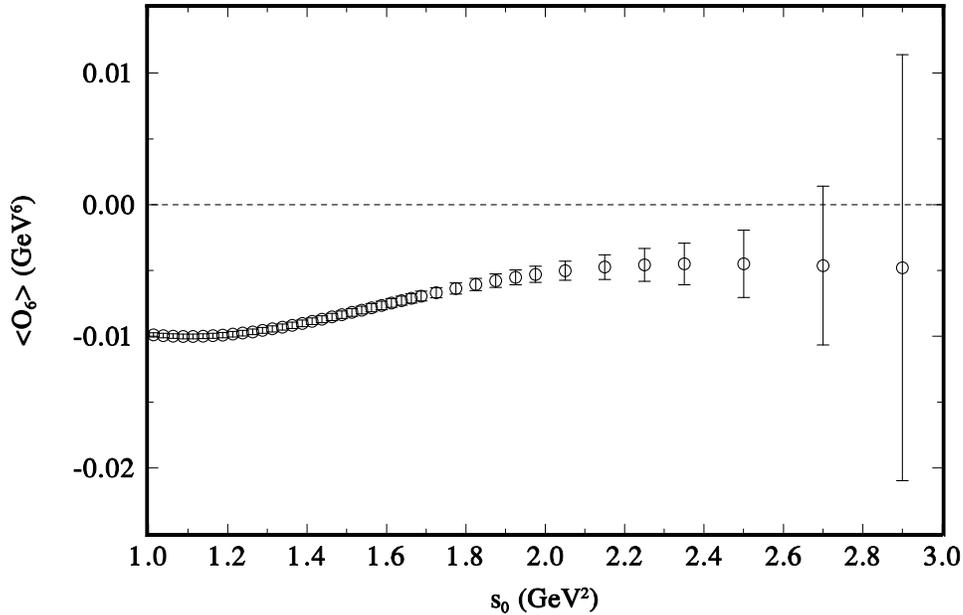}
\caption{
\footnotesize{
The chiral condensate of dimension $d=6$ from the pinched chiral 
sum rule Eq.\ (\ref{Eq:25}). 
}
}
\label{Fig:d=6}
\end{center}
\end{figure} 
%%%%%%%%%%%%%%%%%%%%%%%%%%%%%%%%%%%%%%%

%Fig.7
%%%%%%%%%%%%%%%%%%%%%%%%%%%%%%%%%%%%%%%
\begin{figure}[ht]
\begin{center}
\includegraphics[width=5.2in]{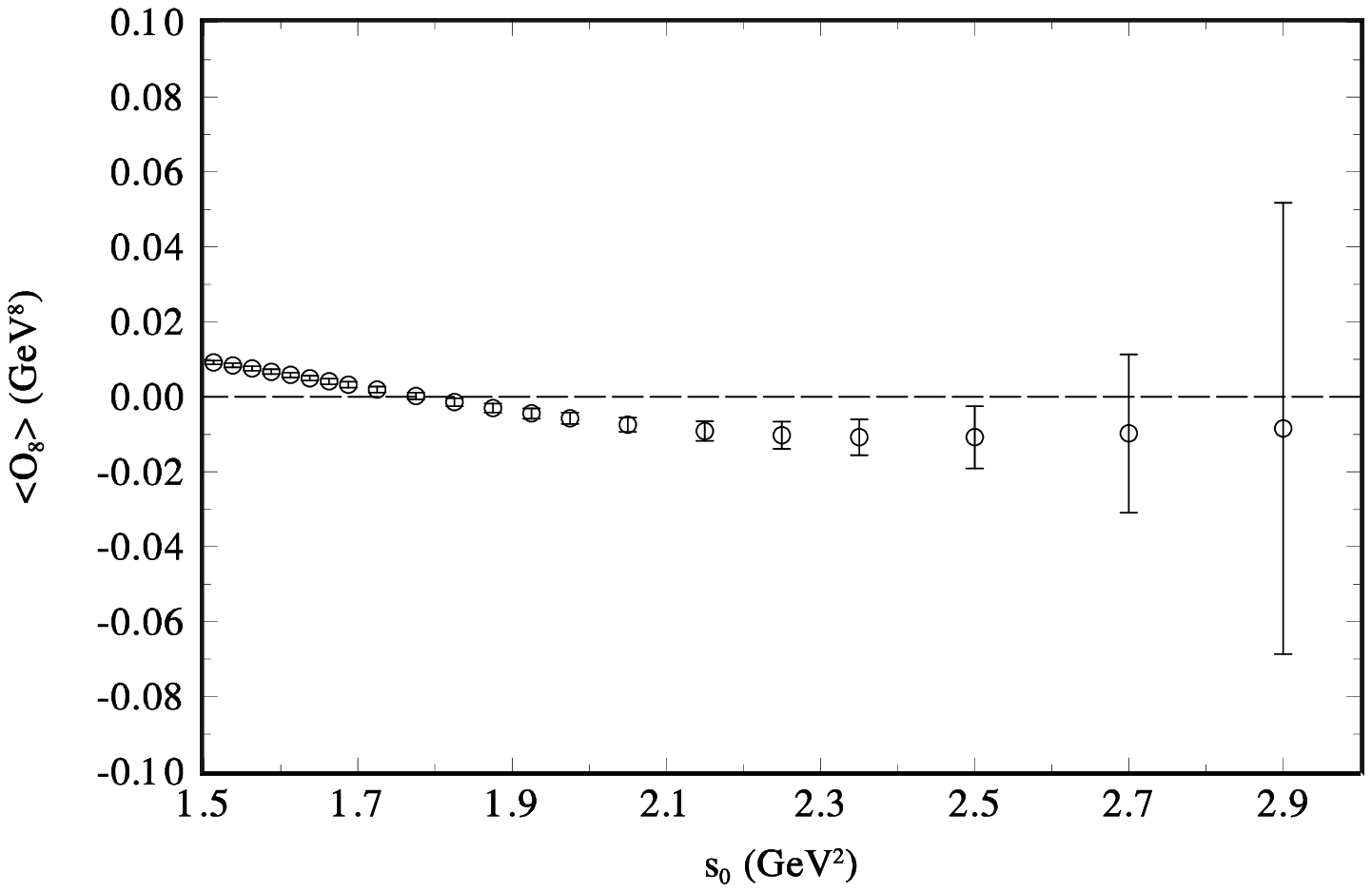}
\caption{
\footnotesize{
The chiral condensate of dimension $d=8$ from the pinched chiral 
sum rule Eq.\ (\ref{Eq:28}). 
}
}
\label{Fig:d=8}
\end{center}
\end{figure} 
%%%%%%%%%%%%%%%%%%%%%%%%%%%%%%%%%%%%%%%

Turning to the chiral condensates, for dimension $d=6$ we use 
the following pinched FESR \cite{CHSR2}
%Eq.30
\begin{equation}
\langle {\cal{O}}_{6}\rangle 
= 
- 2\, f_\pi^2 \, s_0^2 \,+\, s_0^2 \, \int_0^{s_0} ds \, 
\left(1 - \frac{s}{s_0} \right)^2 \; 
\frac{1}{\pi}
[{\mbox{Im}} \Pi_V(s) - {\mbox{Im}} \Pi_A(s)]\;. 
\label{Eq:25}
\end{equation}
The result is shown in Fig.\ \ref{Fig:d=6}. Stability is observed 
for $s_0 \gsim 2$ GeV$^2$. Assuming that DV  are 
not relevant in this kinematic range, we read off the value
%Eq.31
\begin{equation}
\langle {\cal{O}}_{6}\rangle 
= 
- (5.0 \, \pm \, 0.7) \times 10^{-3} \; {\mbox{GeV}}^6.
\label{Eq:26}
\end{equation}
This value  agrees with \cite{CHSR2} obtained from the same sum 
rule, Eq.\ (\ref{Eq:25}), but using the original ALEPH data, i.e. 
$\langle {\cal{O}}_{6}\rangle = - (4.0 \, \pm \, 1.0) \times 10^{-3} 
\; {\mbox{GeV}}^6$. It also agrees  with \cite{L101}, i.e. 
$\langle {\cal{O}}_{6}\rangle = - (4.3 \, \pm \, 0.9) \times 10^{-3} 
\; {\mbox{GeV}}^6$, as well as with \cite{L102} 
$\langle {\cal{O}}_{6}\rangle = - (6.6 \, \pm \, 1.1) \times 10^{-3} 
\; {\mbox{GeV}}^6$. In addition, this value agrees within errors 
with the four-quark condensate in the  vacuum-saturation approximation  
\cite{VS}
%Eq.32
\begin{equation} 
\langle {\cal{O}}_{6}\rangle|_{VS} 
= 
- \frac{64 \, \pi}{9}\, \alpha_s \langle \bar{q} q \rangle^2\, 
\left[1 + \frac{247}{48 \, \pi}\, \alpha_s(s_0) \right] 
\simeq 
- 4.6 \,\times 10^{-3} \,{\mbox{GeV}}^6.
\label{Eq:27}
\end{equation}

Finally, we determine the $d=8$ chiral condensate  from the pinched 
sum rule \cite{CHSR2}
%Eq.33
\begin{equation}
\langle {\cal{O}}_{8}\rangle 
= 
16 f_\pi^2 \, s_0^3 \,- 3\, s_0^4\, \bar{\Pi}(0) 
+
\, s_0^3 \, \int_0^{s_0} \frac{ds}{s} \, 
\left(1 - \frac{s}{s_0} \right)^3 (s + 3\, s_0) \; 
\frac{1}{\pi}
[{\mbox{Im}} \Pi_V(s) - {\mbox{Im}} \Pi_A(s)]\;.
\label{Eq:28}
\end{equation}
The result is shown in Fig.\ \ref{Fig:d=8}, which leads to
%Eq.34
\begin{equation}
\langle {\cal{O}}_{8}\rangle 
= 
- (9.0 \, \pm \, 5.0) \times 10^{-3} \; {\mbox{GeV}}^8 \;,
\label{Eq:29}
\end{equation}
a considerably more accurate value than that of \cite{CHSR2}, 
$\langle {\cal{O}}_{8}\rangle = - (1.0 \, \pm \, 6.0) \times 10^{-3} 
\; {\mbox{GeV}}^8$, as well as that of \cite{L102} 
$\langle {\cal{O}}_{8}\rangle =  (5.0 \, \pm \, 5.0) \times 10^{-3} 
\; {\mbox{GeV}}^8$. The present result does agree within errors with 
that of \cite{L101} $\langle {\cal{O}}_{8}\rangle = - (7.2 \, \pm \, 
4.8) \times 10^{-3} \; {\mbox{GeV}}^8$.

%%%%%%%%%%%%%%%%%%%%%%%%%%%%%%%%%%%%%%%%%%%%%%%%%%%%%%%%%%%%%%%%%%%%%%%%%

\section{Conclusion}

The new ALEPH data base \cite{ALEPH2} has been used together with 
QCD FESR to redetermine a potential dimension $d=2$ term in the OPE, 
as well as the dimension $d=4$ vacuum condensate, i.e. the gluon 
condensate in the chiral limit. The former term is consistent with 
zero, thus confirming expectations \cite{COND3b}, as well as previous 
results \cite{COND4}, while the latter is affected by a larger 
uncertainty than  the result from the original ALEPH data base 
\cite{COND4}. It is important to notice that the current uncertainty 
in the strong coupling (at the scale of the $\tau$-lepton mass) 
dominates over the data errors in the final uncertainty in the 
condensates as obtained from FESR.

The two Weinberg sum rules are saturated by the data at the end of 
the $\tau$-decay spectrum. A simple pinched combination of the 
Weinberg sum rules as well as the Das-Mathur-Okubo sum rule have 
turned out to be amazingly well saturated at much lower center-of-mass 
energies. We consider this as an indication that DV 
are not needed.  We are not asserting, though, that DV do not exist. Instead, we 
interpret this good saturation as suggesting  that our pinched kernels might have
 quenched any potential DV.  However, this conclusion is not universally
  accepted \cite{L101},\cite{L102}. Given the unavoidable need of specific
   models to account for the postulated DV, this issue remains currently an 
   open problem. Similar pinched integration kernels were then used here to determine 
   the chiral correlator at zero momentum, as well as the chiral condensates of
   dimension $d=6$ and $d=8$. In comparison with results using the original ALEPH
    data base, the major changes are 
in the values of the gluon condensate and  of the chiral condensates. 
\\

\bigskip

%%%%%%%%%%%%%%%%%%%%%%%%%%%%%%%%%%%%%%%%%%%%%%%%%%%%%%%%%%%%%%%%%%%%%%%%%

\begin {Large}
{\bf Acknowledgements}
\end{Large}

This work was supported in part by DFG (Germany), and by NRF and 
NITheP (South Africa). One of us (CAD) wishes to thank Andrew 
Hamilton for a discussion on the error analysis.

%%%%%%%%%%%%%%%%%%%%%%%%%%%%%%%%%%%%%%%%%%%%%%%%%%%%%%%%%%%%%%%%%%%%%%%%%

\end{document}